\documentclass[twoside,twocolumn]{article}
\usepackage[super,sort&compress,comma]{natbib} 
\usepackage[version=3]{mhchem}
\usepackage{times,mathptm}
\usepackage{sectsty}
\usepackage{balance} 

\usepackage{graphicx} 
\usepackage{lastpage}
\usepackage[format=plain,singlelinecheck=false,font=small,labelfont=bf,labelsep=space]{caption} 
\usepackage{fancyhdr}
\usepackage{epstopdf}
\begin{document}

\thispagestyle{plain}
\fancypagestyle{plain}{
\renewcommand{\headrulewidth}{1pt}}
\renewcommand{\thefootnote}{\fnsymbol{footnote}}
\renewcommand\footnoterule{\vspace*{1pt}%
\hrule width 3.4in height 0.4pt \vspace*{5pt}} 

\makeatletter 
\renewcommand\@biblabel[1]{#1}
\renewcommand\@makefntext[1]%
{\noindent\makebox[0pt][r]{\@thefnmark\,}#1}
\makeatother 
\renewcommand{\figurename}{\small{Fig.}~}
\sectionfont{\large}
\subsectionfont{\normalsize} 

\fancyfoot{}
\fancyfoot[CO]{\vspace*{-6.5pt}\hspace*{ 11.0cm} \footnotesize{Submitted to Phys. Chem. Chem. Phys.,}}
\fancyfoot[CE]{\vspace*{-6.6pt}\hspace*{-12.7cm} \footnotesize{Submitted to Phys. Chem. Chem. Phys.,}}
\fancyfoot[RO]{\footnotesize{\sffamily{1--\pageref{LastPage} ~\textbar  \hspace{2pt}\thepage}}}
\fancyfoot[LE]{\footnotesize{\sffamily{\thepage~\textbar\hspace{4.6cm} 1--\pageref{LastPage}}}}
\fancyhead{}
\renewcommand{\headrulewidth}{1pt} 
\renewcommand{\footrulewidth}{1pt}
\setlength{\arrayrulewidth}{1pt}
\setlength{\columnsep}{6.5mm}
\setlength\bibsep{1pt}

\twocolumn[
 \begin{@twocolumnfalse}
\noindent\LARGE{\textbf{Long-range Energy Transfer and Ionization in Extended Quantum Systems Driven by Ultrashort Spatially Shaped Laser Pulses}}
\vspace{0.6cm}

\noindent\large{\textbf{Guennaddi K.~Paramonov\textit{$^{a}$},  Andr\'e D.~Bandrauk\textit{$^{b}$}, and Oliver K\"uhn\textit{$^{a}$}$^{\ast}$}}\vspace{0.5cm}


\noindent \textbf{\small{DOI: 10.1039/b000000x}}
\vspace{0.6cm}

\noindent \normalsize{
The processes of ionization and energy transfer in a quantum system composed
 of two distant H atoms with an  initial internuclear separation 
 of 100 atomic units (5.29 nm) have been studied  by the numerical solution of the time-dependent Schr\"odinger equation
 beyond the Born-Oppenheimer approximation. Thereby it has been assumed that only one of the two
 H atoms was excited by temporally and spatially shaped laser pulses
 at various laser carrier frequencies.  The quantum dynamics of the extended H-H system,
 which was taken to be initially either in an unentangled or an entangled ground state,
 has been explored   within a linear three-dimensional model, including  two $z$ coordinates of the electrons
 and the internuclear distance $R$.  An efficient energy transfer from the laser-excited H atom (atom~{\rm A}) 
 to the other H atom (atom~{\rm B})  and the ionization of the latter have been found.
 It has been shown that the physical mechanisms of the energy transfer  as well as of the ionization of atom~{\rm B} are   the Coulomb attraction of the laser driven electron of atom~{\rm A} by
 the proton of atom~{\rm B} and a short-range Coulomb repulsion of the two electrons 
 when their wave functions strongly overlap in the domain of atom~{\rm B}.
}
\vspace{0.5cm}
 \end{@twocolumnfalse}
]



\footnotetext{\textit{$^{a}$~Institut f\"ur Physik, Universit\"at Rostock, D-18051 Rostock, Germany. Fax: +49 381 498 6942; Tel: +49 381 498 6950; E-mail: oliver.kuehn@uni-rostock.de}\\
\textit{$^{b}$~Laboratorie de Chimie Th\'eorique, Facult\'e des Sciences,  Universit\'e de Sherbrooke, Sherbrooke, Qu\'ebec, Canada J1K 2R1}
}



\section{Introduction}
\label{sec:intro}
%
 Long-range energy transfer among distant quantum systems
 has attracted considerable interest recently in  theory and experiment.
 \cite{Cederbaum:97prl.ICD,%
 Marburger:03prl.ICD,%
 Jahnke:04prl.ICD,%
 Morishita:06prl.ICD,%
 Lablanquie:07jcp.ICD,%
 Jahnke:10NaturePhys.ICD,%
 Havermeier:10prl.ICD,%
 Sisourat:10NatPhys.ICD}
Extended quantum systems such as those composed of two atoms  separated by several nanometers are also  referred to as `extreme quantum systems'.
 \cite{Sisourat:10NatPhys.ICD}
 The energy transfer from an excited atom to its neighbor, the so-called interatomic Coulombic decay (ICD), has been first studied
 by Cederbaum {\em et al.} for molecular clusters.
 \cite{Cederbaum:97prl.ICD}
 ICD is well established experimentally  for inner-valence excitation  of many electron systems.
 \cite{Marburger:03prl.ICD,%
 Jahnke:04prl.ICD,%
 Morishita:06prl.ICD,%
 Lablanquie:07jcp.ICD,%
 Jahnke:10NaturePhys.ICD}
 In recent work
 \cite{Havermeier:10prl.ICD},
 ICD was demonstrated experimentally for a helium dimer at interatomic distances
 up to about 12~a.u. Since helium atoms have no inner-valence electrons, 
 a different type of ICD has been suggested for this case which does not require an overlap of the electronic wave functions. \cite{Havermeier:10prl.ICD}

The present work is addressed to an extended quantum system composed of two H atoms, to be referred to as atoms {\rm A} and {\rm B}. The atoms are assumed to be initially separated by an internuclear distance 
 of $R=100$~a.u. which, although chosen somewhat arbitrary, will highlight the particular properties of the laser driven dynamics of such systems.
 The extended H-H system interacts with laser pulses having narrow spatial envelopes,
 such that practically only atom~{\rm A} is excited by the laser field directly,
 while atom~{\rm B} is almost not affected.
 The laser field is  linearly polarized along
 the $z$-axis, and the protons and electrons are assumed to move
 along the polarization direction of the laser  field only.
Three different initial conditions will be considered, which comprises one being of atomic origin where the 
(unentangled) electronic wave function is given by the direct product of the respective atomic wave functions, and two corresponding to a molecular origin where the (entangled) electronic wave function is either symmetrized (singlet) or antisymmetrized (triplet).

In our recent work  \cite{P-K-B:2011pra.H-H.extended}, energy transfer and ionization  in this extended H-H system excited by spatially shaped laser pulses (both narrow and Gaussian type shape) have been studied at the laser carrier frequency of $\omega=1.0$~a.u., corresponding to the ground-state energy of H-H at  $R=100$~a.u.. Both unentangled atomic states and entangled molecular singlet states were used as the initial states in ref. \citenum{P-K-B:2011pra.H-H.extended}. In the present work, we extend this study to a wide range of laser  carrier frequencies,
 including $\omega=0.5$~a.u. corresponding to the ground-state energy of an individual H atom. Further, we will include the case of an initial triplet state. As far  the spatial shape of the laser field is concerned we will consider the case of a narrow shape only, which excites dominantly one atom. Therefore, such a pulse is most suitable for investigating energy transfer and ionization in this extended quantum systems. For a discussion of the effect of a broad spatial envelope we refer to ref. \citenum{P-K-B:2011pra.H-H.extended}.
 
 The relative simplicity of the H-H model used here, i.e.
one-dimensional electrons (coordinates $z_{1}$ and $z_{2}$) and protons (coordinate $R$),
 makes it possible to treat  the long-range electronic motion explicitly together with the nuclear motion
 and to reveal the role played by the overlap of the electronic wave functions and by the electron-electron Coulombic repulsion. However, effects similar to those presented below can be anticipated to occur
 in more complex extended quantum systems as well.  Such systems can be in the bound state, as a helium dimer
 \cite{Havermeier:10prl.ICD}  for example, or represent spatially well separated atoms as those studied in the present work.

 The paper is organized as follows. The model, equations of motion and techniques 
 are described in section \ref{sec:methods}. 
 Results obtained for energy transfer and ionization of H-H in a wide range 
 of carrier frequencies  of the applied temporally and spatially shaped laser pulses 
 and the laser-driven dynamics of H-H at several carrier frequencies
 are presented in section \ref{sec:results}. 
 The results obtained are summarized  in the concluding section \ref{sec:conc}.
%
\section{Theoretical Model and Computational Methods}
\label{sec:methods}
%
\subsection{Time-dependent Schr\"odinger Equation}
Within the linear three-dimensional model used in this paper to describe the H-H system
 excited by temporally and spatially shaped laser fields,
 the total Hamiltonian $\hat H_{\rm T}$ 
 consists of two parts,
 \begin{equation}
 \hat H_{\rm T}(R,z_{1},z_{2},t) = \hat H_{\rm S}(R,z_{1},z_{2}) + \hat H_{\rm SF}(z_{1},z_{2},t),
  \label{E-2}
 \end{equation}
 where $\hat H_{\rm S}(R,z_{1},z_{2})$ represents the H-H system and
 $\hat H_{\rm SF}(z_{1},z_{2},t)$ describes its interaction 
 with the laser field. 
 The applied laser field is linearly polarized along
 the $z$-axis, the nuclear and the electronic motion are restricted
 to the polarization direction of the laser electric field.
 Accordingly, the two $z$ coordinates of electrons, $z_{1}$ and $z_{2}$,
 measured with respect to the nuclear center of mass,
 are treated explicitly together with the internuclear distance $R$. 
 A similar model has been used previously, e.g. in ref.  \citenum{Bandr:08prl.H2d3},
 for the H$_{2}$ molecule. 

 The total non-Born-Oppenheimer Hamiltonian of the H-H system  (employing atomic units) reads
 \begin{displaymath}
 \hat H_{\rm S}(R,z_{1},z_{2}) =
 -\frac{1}{m_{\rm p}}\frac{\partial^{2}}{\partial R^{2}} 
 + V_{\rm p p}(R) 
 \end{displaymath}
 \begin{equation}
 + \sum _{k=1}^{2} \biggl[-\frac{1}{2 \mu_{\rm e}}\frac{\partial^{2}}{\partial z_{k}^{2}} 
  + V_{\rm e p}(z_{k},R)\biggr] + V_{\rm e e}(z_{1},z_{2}),
  \label{E-3}
 \end{equation}
 where $m_{\rm p}$ is the proton mass, $\mu_{\rm e}=2m_{\rm p}/(2m_{\rm p}+1)$
 is the electron reduced mass, and non-diagonal mass-polarization terms
 are neglected.
 The Coulomb potentials in eqn~(\ref{E-3}) are
 \begin{displaymath}
 V_{\rm p p}(R)=\frac{1}{R}, \, \, \, \, \, \, \, 
 V_{\rm e e}(z_{1},z_{2})=\frac{1}{\sqrt{(z_{1}-z_{2})^{2}+\alpha}},
 \end{displaymath}
 \begin{equation}
 V_{\rm e p}(z_{k},R)=-\frac{1}{\sqrt{(z_{k}-R/2)^{2}+\beta}}-\frac{1}{\sqrt{(z_{k}+R/2)^{2}+\beta}},
  \label{E-4}
 \end{equation}
 where $k=1,2$, and the regularization parameters,  $\alpha=0.1\times 10^{-3}$ and $ \beta=1.995$, 
 have been chosen   such as to reproduce the ground-state (GS) energy    of the H-H system at $R=100$~a.u.
 (${\rm E}_{\rm H-H}^{\rm GS}=-1.0$~a.u.).

 The interaction of the H-H system with the laser field, $\hat H_{\rm SF}(z_{1},z_{2},t)$,
 is treated semiclassically  within the electric dipole approximation via the vector potential,
 \begin{equation}
 \hat H_{\rm SF}(z_{1},z_{2},t) = -\frac{1}{c}\frac{\partial A(t)}{\partial t}
  (1+\gamma)
   \sum _{k=1}^{2} F(z_{k}) z_{k}, 
  \label{E-6}
 \end{equation}
 where $\gamma=(1+2m_{\rm p})^{-1}$,  
 $A(t)$ is the vector potential, $c$ is the speed of light,
 and $F(z)$ is the spatial-shape function, or spatial envelope, of the laser field.
 The following, sin$^{2}$-type, narrow spatial envelope of the laser pulse 
 acting practically only on atom~{\rm A} was used for 
 $z=z_{1}$ and $z=z_{2}$:
 \begin{equation}
  F(z)=\sin^{2}\biggl[\frac{\pi(z-z_{a})}{z_{b}-z_{a}}\biggr], \; \; \; z_{a} \leq z \leq z_{b}, 
  \label{E-N}
 \end{equation}
 where $z_{a}=-60$~a.u., $z_{b}=-40$~a.u., and $F(z)=0$ otherwise.

 The vector potential $A(t)$ in eqn~(\ref{E-6}) is chosen as follows:
 \begin{equation}
 A(t)=\frac{c}{\omega}{\cal E}_{0}\sin^{2}(\pi t/t_{p})\cos(\omega t + \phi),
  \label{E-7}
 \end{equation}
 where ${\cal E}_{0}$ is the amplitude, 
 $t_{p}$ is the pulse duration,
 $\omega $ is the laser carrier frequency,
 and $\phi$ is the carrier-envelope phase. 
 The definition of the system-field interaction  via the vector potential by eqn~(\ref{E-6}), 
 suggested in  ref. \citenum{Bandrauk:02pra.VPA},  ensures that the electric field
 has a vanishing direct-current   component  and satisfies Maxwell's equations in the propagation region.

 It is suitable to define, on the basis of  eqns~(\ref{E-6}) and (\ref{E-7}),  the local effective-field amplitude for atom~{\rm A}
 as follows:
 \begin{equation}
 {\cal E}_{0}^{\rm{\rm A}}={\cal E}_{0} F(z_{\rm A}), \; \; \;
  \label{E-Eampl}
 \end{equation}
 where  $z_{\rm A}=-50$~a.u. (atom~{\rm B} is located initially at $z_{\rm B}=50$~a.u.).
 The respective time-dependent electric field acting on atom {\rm A} reads
 \begin{displaymath}
 {\cal E}^{{\rm A}}(t)={\cal E}_{0}^{{\rm A}}[\sin^{2}(\pi t/t_{p})\sin(\omega t + \phi)
 \end{displaymath}
 \begin{equation}
 -\frac{\pi}{\omega t_{p}}\sin(2\pi t/t_{p})\cos(\omega t + \phi)].
  \label{E-Etime}
 \end{equation}
 The first term in eqn~(\ref{E-Etime}) corresponds to a laser pulse  with a sin$^{2}$-type temporal envelope,
 and the second term appears  due to the finite pulse duration.
 \cite{Bandrauk:02pra.VPA,%
 Doslic:06pra.VPA}

 The 3D time-dependent Schr\"odinger equation,
 \begin{equation}
 i\frac{\partial}{\partial t} \Psi = 
 \bigl[\hat H_{\rm S}(R,z_{1},z_{2})+\hat H_{\rm SF}(z_{1},z_{2},t)\bigr] \Psi ,
  \label{TDSE}
 \end{equation}
 has been solved numerically with the propagation technique adapted from refs.
 \citenum{Paramon:05cpl.HH-HD} and \citenum{ Paramon:07cp.HH-HD-Muon}
 for both electron and proton motion.  In particular, calculations for the electron motion have been performed
 by using 200-point non-equidistant grids  for the Hermite polynomials and corresponding weights for the numerical
 integration on the interval $(-\infty, \infty)$ for the $z_{1}$ and $z_{2}$  coordinates. For the nuclear coordinate $R$, a 256-point equidistant grid
 has been used on the interval $[75$~a.u., $125$~a.u.$]$.   The time-step of the propagation was $\Delta t = 0.021$~a.u..
 
  The wave function of the initial state has been obtained by the numerical
 propagation of the equation of motion, eqn (\ref{TDSE}), in the imaginary time,
 with the laser field being switched off (${\cal E}_{0}=0$). Three cases will be considered, which correspond to different preparation conditions and are either of atomic or molecular origin (see next subsections).
Note before proceeding that the case of unentangled atomic states is more likely
 to be realized experimentally. For example, at a gas pressure of 1~atm.,
 the interatomic distance is about 100~a.u.. 
 In contrast entangled molecular states
 require additional steps such as dissociation of H$_{2}$ molecules or generation of entanglement between two individual H atoms. 
 
The wave function analysis will be performed on the basis of the electron probabilities $P(z_{1})$ and $P(z_{2})$ defined as follows:
 \begin{equation}
 P(z_{1/2})=\int dR \int dz_{2/1} |\Psi (R,z_{1},z_{2})|^{2} 
  \label{E-5A}
 \end{equation}
 for electron $e_{1/2}$ initially belonging to atom {\rm A/B} with proton $p_{\rm A/B}$.
 These electron probabilities give the overall probability to find an electron
 at a specified point of the $z$-axis at any position of the other electron 
 and at any internuclear distance. 

\subsection{Product Initial State}
In this case, which is termed to be of \emph{atomic origin} (two distant H atoms), the spatial part of the initial unentangled ground-state wave function of H-H  used in the imaginary time propagation was defined by the non-symmetrized 
 Heitler-London wave function as follows:
 \begin{equation}
 \Psi(R,z_{1},z_{2},t=0)=\Psi_{1S_{\rm A}}(z_{1}) \Psi_{1S_{\rm B}}(z_{2}) \Psi_{\rm G}(R),
  \label{ini-DP}
 \end{equation}
 where $\Psi_{1S_{\rm A}}(z_{1})=e^{-|z_{1}-z_{\rm A}|}$  at  $z_{\rm A}=-50$~a.u.,
 $\Psi_{1S_{\rm B}}(z_{2})=e^{-|z_{2}-z_{\rm B}|}$ at $z_{\rm B}=50$~a.u.,
 and $\Psi_{\rm G}(R)$ is a Gaussian function of unit width and centered at $R=100$~a.u..

  Upon excitation of the extended H-H system by the laser field, 
 the electronic wave functions of 
 its atomic {\rm A} and {\rm B} parts may overlap.
 In order to study the energy transfer from atom~{\rm A} to atom~{\rm B},
 the respective `atomic' energies, $E_{\rm A}(t)$ and $E_{\rm B}(t)$, 
 have been defined  
 on the basis of eqns.~(\ref{E-3}) and (\ref{E-4}) 
 as follows: 
 \begin{displaymath}
 E_{\rm A/B}(t) = \biggl< \Psi (t) \bigg|
 -\frac{1}{2m_{\rm p}}\frac{\partial^{2}}{\partial R^{2}} 
 + \frac{1}{2} V_{\rm p p}(R) 
  + \frac{1}{2} V_{\rm e e}(z_{1},z_{2})
 \end{displaymath}
 \begin{equation}
  -\frac{1}{2 \mu_{\rm e}}\frac{\partial^{2}}{\partial z_{1/2}^{2}} 
  - \sum _{k=1}^{2} \frac{1}{\sqrt{(z_{k}\pm R/2)^{2}+\beta}}
  \bigg| \Psi (t) \biggr> \, .
  \label{E-EnA-DP}
 \end{equation}
 It is seen 
 that the kinetic energy of nuclear motion and the potential energies 
 of the proton-proton and the electron-electron interaction are assumed
 to be equally shared between atoms~{\rm A} and {\rm B}. 
 The kinetic energy of electron $e_{1}$ 
 and the energy of Coulombic interaction of both electrons $e_{1}$ and $e_{2}$
 with proton $p_{\rm A}$
 are entirely assigned to atom {\rm A}.
 Similarly, the kinetic energy of electron $e_{2}$
 and the energy of Coulombic interaction of both electrons
 with proton $p_{\rm B}$
 are entirely assigned to atom {\rm B}.
 The sum of `atomic' energies gives the correct total energy of the entire H-H system.

 The ionization probabilities for atoms {\rm A} and {\rm B} 
 have been calculated 
 from the time- and space-integrated outgoing fluxes, 
 separately, for the positive and the negative directions
 of the $z_{1}$ and $z_{2}$ axes 
 at $z_{1,2}=\pm 91$~a.u..
 We calculated four ionization probabilities:
 ${\rm I}_{\rm A}(z_{1}=-91\,{\rm a.u.})$ and ${\rm I}_{\rm A}(z_{2}=-91\,{\rm a.u.})$
 for atom {\rm A}, and
 ${\rm I}_{\rm B}(z_{1}=91\,{\rm a.u.})$ and ${\rm I}_{\rm B}(z_{2}=91\,{\rm a.u.})$
 for atom {\rm B}.
 At the outer ends of the $z$-grids, absorbing boundaries have been
 provided by imaginary smooth optical potentials adapted from 
 that designed in ref.
 \citenum{Rabitz:94.jcp.HF}. 
 Similar optical potentials have  also been included for the $R$-axis,
 but in practice the wave-packet never approached the outer ends
 of the $R$-grid.
 \subsection{Entangled Initial States} 
  The initial states of molecular H-H systems, singlet and triplet,
 are entangled by spin exchange and represented by the Heitler-London
 symmetrized and antisymmetrized products of atomic wave functions, respectively.
 The spatial parts of the initial entangled wave functions
 for singlet (S) and triplet (T) molecular electronic states are 
 given by
 \begin{displaymath}
 \Psi(R,z_{1},z_{2},t=0)_{\rm S/T}=[\Psi_{1S_{\rm A}}(z_{1}) \Psi_{1S_{\rm B}}(z_{2})
 \end{displaymath}
 \begin{equation}
 \pm \Psi_{1S_{\rm B}}(z_{1}) \Psi_{1S_{\rm A}}(z_{2})] \Psi_{\rm G}(R),
  \label{ini-ST}
 \end{equation}
 where $\Psi_{1S_{\rm A,B}}(z_{1,2})$ are defined by eqn~(10)
 and $\Psi_{\rm G}(R)$ is a proton Gaussian function centered at $R=100$~a.u..
 Imaginary time propagations have been performed 
 with the initial wave functions (\ref{ini-ST})
 and the system Hamiltonian of eqn~(2). 
 The energies of the singlet and triplet molecular states
 of the extended H-H system at the large internuclear separation of $R=100$~a.u. 
 are equal to each other and to the energy of the unentangled atomic initial state.

 For the entangled initial states (\ref{ini-ST}),
 the `atomic' energies
 $E_{\rm A}(t)$ and $E_{\rm B}(t)$ are defined on the basis
 of eqns.~(2) and (3) as follows:
 \begin{displaymath}
 E_{\rm A/B}(t) = \biggl< \Psi (t) \bigg|
 -\frac{1}{2m_{\rm p}}\frac{\partial^{2}}{\partial R^{2}}
  - \frac{1}{2}\sum _{k=1}^{2} \frac{1}{2 \mu_{\rm e}}\frac{\partial^{2}}{\partial z_{k}^{2}}
 + \frac{1}{2} V_{\rm p p}(R)
 \end{displaymath}
 \begin{equation}
  + \frac{1}{2} V_{\rm e e}(z_{1},z_{2})
 - \sum _{k=1}^{2} \frac{1}{\sqrt{(z_{k} \pm R/2)^{2}+\beta}}
  \bigg| \Psi (t) \biggr>
  \label{E-EnAB-TS}
 \end{equation}
 It is seen
 that the kinetic energies of nuclear and electronic motion
 as well as the potential energies of
 the proton-proton and the electron-electron Coulombic interaction are assumed
 to be equally shared between atoms {\rm A} and {\rm B}.
 The energy of Coulombic interaction of both electrons with
 proton $p_{\rm A}$ is assigned to atom~{\rm A}, while
 the energy of Coulombic interaction of both electrons with
 proton $p_{\rm B}$ is assigned to atom~{\rm B}.
 The sum of these `atomic' energies
 always gives the correct total energy of the entire H-H system.
 The choice of the electron kinetic energies in eqn~(\ref{E-EnAB-TS})
 is different from that used in eqn~(12) 
 and corresponds to the initial electron probabilities $P(z_{1})$ and $P(z_{2})$ 
 in the entangled molecular states. 
 Indeed, both electrons $e_{1}$ and $e_{2}$ are localized with the 50\%
 probability in the vicinity of proton $p_{\rm A}$ of atom~{\rm A} and when the extended
 H-H system is excited  by the narrowly shaped laser pulse of Fig.~1(b),
 both electrons give rise to the `atomic' energy $E_{\rm A}(t)$ with the same probability.

 Our numerical simulations have shown that
 the ionization probabilities of electrons $e_{1}$ and $e_{1}$
 excited from the entangled initial states, singlet and triplet,
 are almost identical to each other both for the positive and the negative directions
 of the $z$-axes:
 ${\rm I}_{\rm A}(z_{1}=-91\,{\rm a.u.})={\rm I}_{\rm A}(z_{2}=-91\,{\rm a.u.})$
 and
 ${\rm I}_{\rm B}(z_{1}=91\,{\rm a.u.})={\rm I}_{\rm B}(z_{2}=91\,{\rm a.u.})$.
 This is the consequence of the symmetry of the entangled initial wave functions (cf. ref. \citenum{P-K-B:2011pra.H-H.extended}).
 It is therefore reasonable to define the total ionization probabilities
 for atoms {\rm A} and {\rm B} as follows:
 \begin{displaymath}
 {\rm I}_{\rm A}^{\rm Total}={\rm I}_{\rm A}(z_{1}=-91\,{\rm a.u.})
 +{\rm I}_{\rm A}(z_{2}=-91\,{\rm a.u.}),
 \end{displaymath}
 \begin{equation}
 {\rm I}_{\rm B}^{\rm Total}={\rm I}_{\rm B}(z_{1}=91\,{\rm a.u.})
 +{\rm I}_{\rm B}(z_{2}=91\,{\rm a.u.}).
  \label{E-Flexes}
 \end{equation}
%
 \section{Numerical Results} 
 \label{sec:results}
\subsection{Product Initial State}
\begin{figure}[t]
\begin{center}
 \includegraphics*[width=16pc]{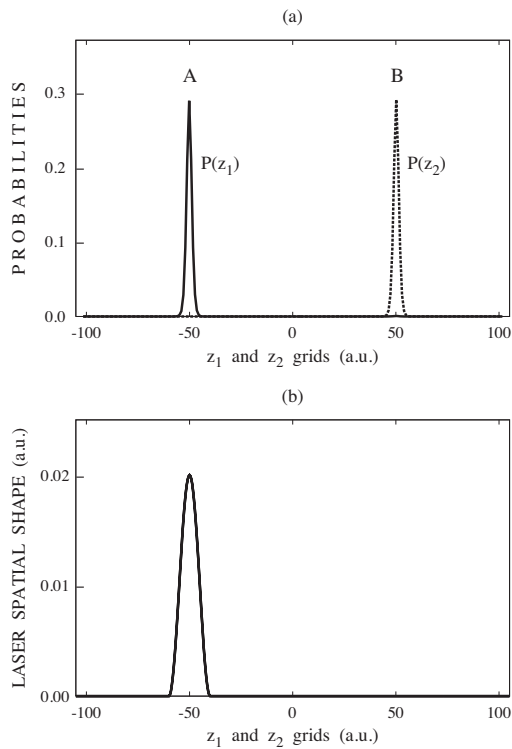}
\end{center}
 \caption{The unentangled initial state 
 of the extended H-H system and a narrow spatial shape of the applied laser pulse.
 (a) - electron probabilities $P(z_{1})$ and $P(z_{2})$ 
 defined by eqn~(\ref{E-5A});
 (b) - a narrow spatial envelope of the laser pulse, defined by eqn~(\ref{E-N}), 
 which dominantly excites only electron $e_{1}$ in the domain of atom~{\rm A}.}
 \label{fig1}
 \end{figure}

The probability distribution for the initial state of the extended H-H system representing
 two distant H atoms, {\rm A} and {\rm B}, is shown in Fig.~\ref{fig1}(a). 
  The narrow spatial envelope of the applied laser pulses
 is presented in Fig.~\ref{fig1}(b) to illustrate the local effective-field
 strengths acting on atoms {\rm A} and {\rm B}.
 We clearly see from Fig.~\ref{fig1} that the narrowly shaped laser pulse
excites practically only electron $e_{1}$ (coordinate $z_{1}$)
 in the domain of atom~{\rm A},
 because the probability to find electron $e_{2}$ (coordinate $z_{2}$)
 at $z_{2}=-50$~a.u. is extremely small. 

 The electric field of the laser pulses is defined by eqns~(\ref{E-Eampl}) and (\ref{E-Etime}),
 where we choose the amplitude ${\cal E}_{0}=0.02$~a.u. (corresponding to an intensity of
 $I_{0}=1.4\times 10^{13}$~W/cm$^{2}$) and the pulse duration $t_{p}=5$~fs.
 The laser carrier frequency $\omega$ is varied in a wide range covering, in particular,
 $\omega=0.5$~a.u. (corresponding to the ground-state energy of an H atom) and
 $\omega=1.0$~a.u. (corresponding to the ground-state energy of the extended H-H system).
In passing we note that with the laser pulse duration being $t_{p}=5$~fs the number of optical cycles during the pulse,
 $N_{c}=\omega t_{p}/2\pi$, is about 33 at $\omega=1.0$~a.u., and therefore, 
 the carrier envelope phase (CEP) of
 the laser field, $\phi$, is not important.
 In contrast, at $\omega<0.5$~a.u. the number of optical cycles is $N_{c}<15$,
 and CEP may play an important role.
 \cite{Bandrauk:2002.PRA.HatomNc15}
However, postponing a detailed study of the role of CEP-effects,
 in the present work we set the carrier phase equal to zero 
 [$\phi=0$ in eqns.~(\ref{E-7}) and (\ref{E-Etime})].

\begin{figure}[t]
 \begin{center}
\includegraphics*[width=16pc]{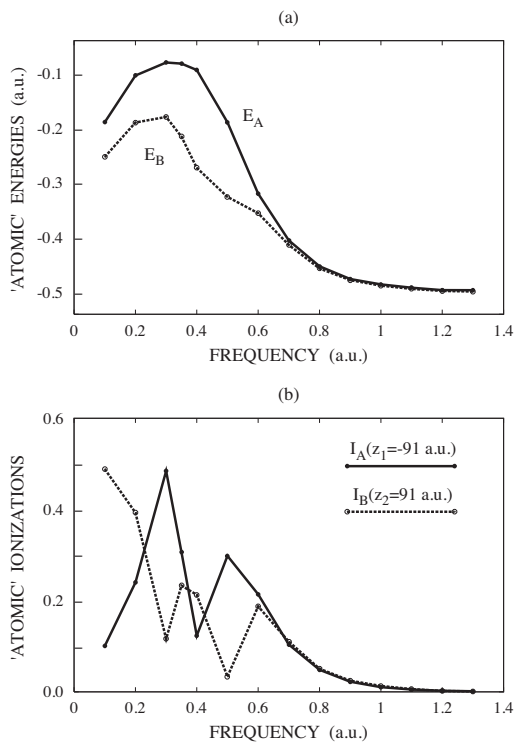}
\end{center} 
\caption{Excitation of the extended H-H system
 by laser pulses with the narrow spatial envelope of eqn~(\ref{E-N})
 at various laser carrier frequencies: 
 `atomic' energies and `atomic' ionizations at $t=10$~fs 
 (5~fs excitation by the laser pulse followed by 5~fs free evolution).
 (a) - `atomic' energies $E_{\rm A}$ and $E_{\rm B}$ 
 defined by eqn~(\ref{E-EnA-DP}); 
 (b) - `atomic' ionizations in positive and negative directions of the $z$-axis. Note that
 the ionization probabilities ${\rm I}_{\rm A}(z_{1}=91\,{\rm a.u.})$ and
 ${\rm I}_{\rm B}(z_{2}=-91\,{\rm a.u.})$ are of the order of 10$^{-4}$ 
 and not included here.}
 \label{fig2}
 \end{figure}

The `atomic' energies $E_{\rm A}$ and $E_{\rm B}$, eqn~(\ref{E-EnA-DP}),
 and `atomic' ionizations ${\rm I}_{\rm A}(z_{1}=-91\,{\rm a.u.})$ and
 ${\rm I}_{\rm B}(z_{2}=91\,{\rm a.u.})$ as obtained from the numerical solution of the time-dependent Schr\"odinger equation for various laser frequencies are shown in Fig. \ref{fig2}. 
 The `atomic' energies and ionizations, presented in Fig.~\ref{fig2}, are calculated
 at $t=10$~fs, which corresponds to the excitation of H-H by a 5~fs laser pulse
 and a 5~fs free evolution of excited H-H after the end of the laser pulse. Notice that in general a small/large `atomic' energy, $E_{\rm B}$, and ionization, ${\rm I}_{\rm B}$, of 
 atom~{\rm B} for an excitation of  atom~{\rm A}
 by the laser pulse with a narrow spatial envelope [Fig.~\ref{fig1}(b)] would evidence a weak/strong energy transfer from {\rm A} to {\rm B}.

 The following observations can be made from Figs.~\ref{fig2}(a) and (b):

 (i) The frequency dependence of both `atomic' energies [Fig.~\ref{fig2}(a)]
 and, especially,
 `atomic' ionizations [Fig.~\ref{fig2}(b)]
 can be divided into two domains: 
 a low-frequency domain ($\omega \leq 0.6$~a.u.) with pronounced changes
 and a high-frequency domain ($\omega > 0.6$~a.u.) with smooth changes. 

 (ii) In the low-frequency domain, including the resonant frequency $\omega = 0.5$~a.u.,
 which corresponds to the one-photon resonant excitation of an isolated H atom
 from its ground state, the 
 `atomic' energies $E_{\rm A}$ and $E_{\rm B}$ [see Fig.~\ref{fig2}(a)]
 first increase with $\omega$ to reach their maxima at $\omega = 0.3$~a.u..
 At larger frequencies, $\omega \geq 0.3$~a.u., $E_{\rm A}$ decreases more rapid than $E_{\rm B}$ such that their
 numerical values approach each other towards the end of this interval. Overall, however,
the  `atomic' energy  $E_{\rm B}$ is substantially smaller than $E_{\rm A}$ in the low-frequency domain.
 
 (iii) In the low-frequency domain ($\omega \leq 0.6$~a.u.), 
 `atomic' ionizations, 
 ${\rm I}_{\rm A}(z_{1}=-91\,{\rm a.u.})$ for atom~{\rm A} 
 and ${\rm I}_{\rm B}(z_{2}=91\,{\rm a.u.})$ for atom~{\rm B},
 demonstrate a very sharp and irregular dependence on 
 the laser carrier frequency $\omega $ [Fig.~\ref{fig2}(b)].
 Indeed, at $\omega = 0.1$ and 0.2~a.u., the ionization probability of
 atom~{\rm B} is large compared with that
 of atom~{\rm A}, implying an efficient transfer of  electronic energy
 from atom~{\rm A} to atom~{\rm B}.
 In contrast, at $\omega = 0.3$~a.u., the ionization probability ${\rm I}_{\rm A}$
 is more than four times larger than ${\rm I}_{\rm B}$.
 At $\omega = 0.4$~a.u. the ionization of atom~{\rm B} dominates again, 
 while at $\omega = 0.5$~a.u. ionization of atom~{\rm A} is almost
 nine times larger than ionization of atom~{\rm B}, implying that 
 energy transfer from atom~{\rm A} to atom~{\rm B} is not efficient
 at a resonant excitation of atom~{\rm A} from its ground state.

 (iv) Finally, in the high-frequency domain ($\omega > 0.6$~a.u.),
 both `atomic' energies and `atomic' ionizations decrease smoothly
 when $\omega$ increases, with $E_{\rm A}$ being close to $E_{\rm B}$,
 and ${\rm I}_{\rm A}$ being close to ${\rm I}_{\rm B}$.
Below  we shall pay attention to the low-frequency domain
 ($\omega \leq 0.6$~a.u.) only, while the behavior in the high-frequency domain
 will be addressed elsewhere.
 
 \begin{figure}[t]
\begin{center}
 \includegraphics*[width=15pc]{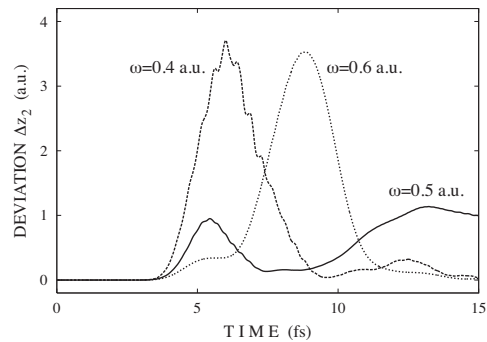}
\end{center}
 \caption{Time-dependent deviation $\Delta z_{2}(t)$ of the coordinate expectation value 
 of electron $e_{2}$ of atom~{\rm B} [Eq.~(\ref{E-Zdeviations})]
 at three laser carrier frequencies, $\omega =$0.4, 0.5 and 0.6~a.u.,
 of the 5~fs laser pulse with the narrow spatial envelope of Eq.~(\ref{E-N})
 which excites only electron $e_{1}$ of atom~{\rm A} of the extended H-H system.}
 \label{fig3}
 \end{figure}

 \begin{figure*}[t]
\begin{center}
 \includegraphics*[width=28pc]{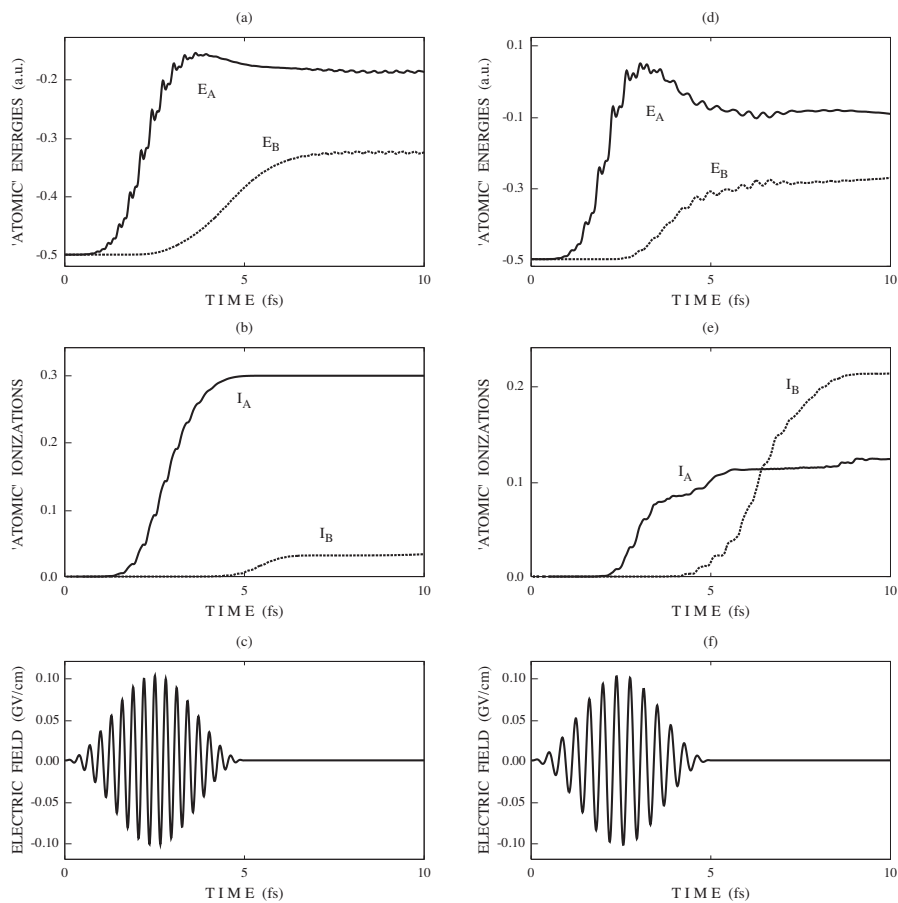}
\end{center}
 \caption{Excitation of the extended H-H system
 by the laser pulse with the narrow spatial envelope of eqn~(\ref{E-N})
 acting only on atom~{\rm A} in the vicinity of $z_{\rm A}=-50$~a.u.:
 an {\it intuitive} case (left panel, $\omega =$0.5~a.u.) and
 a {\it counter-intuitive} case (right panel, $\omega =$0.4~a.u.).
 (a) and (d) - `atomic' energies $E_{\rm A}$ and $E_{\rm B}$;
 (b) and (e) - `atomic' ionizations, 
 ${\rm I}_{\rm A}(z_{1}=-91\,{\rm a.u.})$ for atom~{\rm A} 
 and ${\rm I}_{\rm B}(z_{2}=91\,{\rm a.u.})$ for atom~{\rm B}
 [ionization probabilities ${\rm I}_{\rm A}(z_{1}=91\,{\rm a.u.})$ and
 ${\rm I}_{\rm B}(z_{2}=-91\,{\rm a.u.})$ are of the order of 10$^{-4}$ 
 and are not shown];
 (c) and (f) - laser fields acting on atom~{\rm A}.} 
 \label{fig4}
 \end{figure*}

The most striking observation in the low-frequency domain is that while  the `atomic' energies behave quite smoothly the ionization probabilities show rapid changes in particular when sweeping through the isolated atom resonance at $\omega=$0.5 a.u.
This behavior is reminiscent of the oscillatory dependence of the ionization probability reported for single H-atoms, although for somewhat stronger and shorter laser fields in ref. \citenum{prasad10:055302}.
In the present case the reason of the behavior of ionization probability, e.g.  ${\rm I}_{\rm B}$,
 on the laser carrier frequency $\omega$ can be understood from the time-dependent
 deviation of the coordinate expectation value for electron $e_{2}$ of atom~{\rm B} from its initial position at $t=0$, i.e.
 \begin{equation}
 \Delta z_{2}(t)=\langle z_{2}(t) \rangle-\langle z_{2}(t=0) \rangle.
  \label{E-Zdeviations}
 \end{equation}
 The time-dependent deviation $\Delta z_{2}(t)$ is presented in Fig.~\ref{fig3}
 on the timescale of 15~fs for three laser carrier frequencies:
 $\omega =$0.4, 0.5 and 0.6~a.u..
 From Fig.~\ref{fig3} we see that the dynamics of deviation $\Delta z_{2}(t)$
 at $\omega =$0.4 and 0.6~a.u. is rather similar: 
 in both cases electron $e_{2}$ of atom~{\rm B} gets a momentum in the positive direction
 of the $z_{2}$-axis due to the Coulombic interaction with
 the laser-driven electron $e_{1}$ coming from atom~{\rm A}, which results eventually
 in the ionization of atom~{\rm B} in the positive direction of the $z_{2}$-axis.
 Note that a similar single-peak dynamics of electron $e_{2}$ deviation $\Delta z_{2}(t)$ 
 occurs at all other laser carrier frequencies
 studied in this work (i.e. 0.1~a.u. $\leq \omega \leq $1.3~a.u.), 
 except for the resonant frequency $\omega =$0.5~a.u., when the Coulombic interaction
 of the two electrons is comparatively weak albeit durable (see Fig.~\ref{fig3}, solid line). 
 
 The time delay of the maxima for $\omega =$0.4 and 0.6 a.u. observed in Fig. \ref{fig3} finds a simple explanation in classical terms. With a kinetic energy of 0.1 a.u., which is available to the electron $e_1$ after excitation with $\omega=0.6$ a.u. it can reach atom B within $\sim$5.5 fs. Excitation with $\omega=0.4$ a.u. requires a two photon transition for ionization and therefore the initial kinetic energy will be about 0.3 a.u. so that it takes only $\sim$3 fs for electron $e_1$ to reach atom B. The delay between the maxima of $\Delta z_{2}(t)$ in Fig. \ref{fig3} nicely agrees with this simple estimate. The resonant case $\omega=0.5$ a.u. is special since the created wave packet will be of Rydberg character. Such Rydberg wave packets have recently been studied for H atoms in ref. \citenum{ishikawa10:999}. There it was found that the maximum of the wave packet reaches $\sim$150 a.u. on a time scale as long as 24 fs. Of course, in the present case this wave packet is subject to the Coulombic interaction with atom B and, therefore, it is strongly disturbed as compared with the single H-atom case.

 \begin{figure*}[t]
\begin{center}
 \includegraphics*[width=0.8\textwidth]{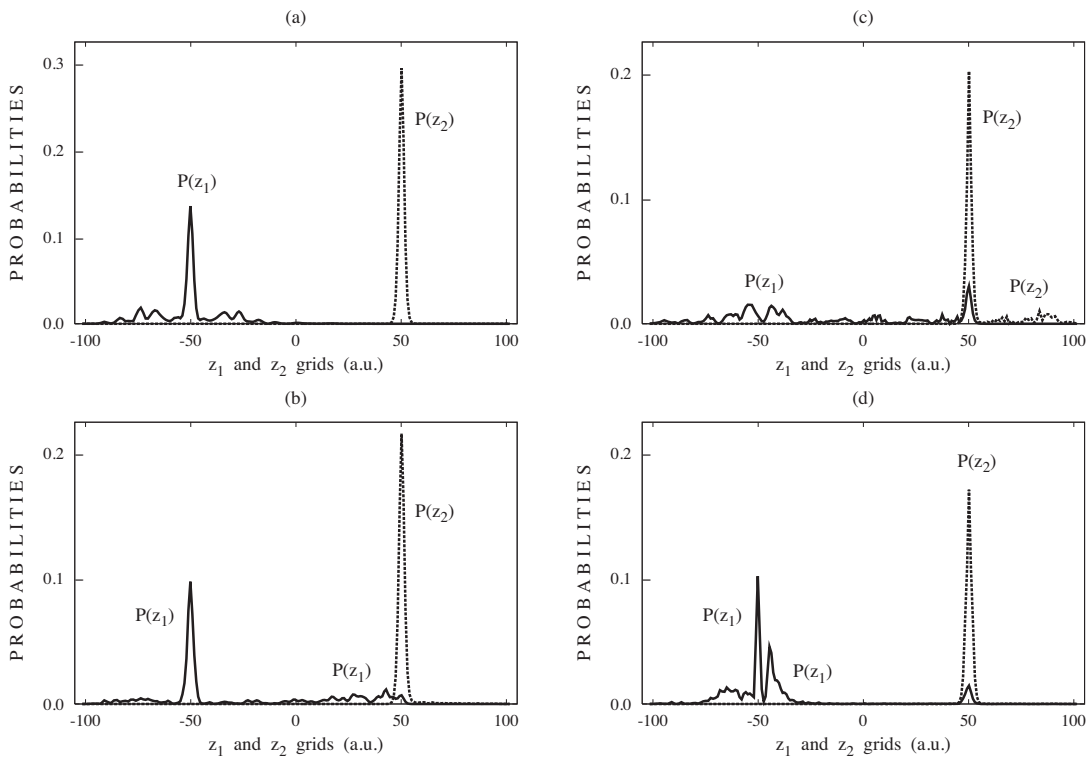}
\end{center}
 \caption{Excitation of the extended H-H system
 by a laser pulse with the narrow spatial envelope of eqn~(\ref{E-N})
 acting only on electron $e_{1}$ (coordinate $z_{1}$) of atom~{\rm A}:
 electron probabilities $P(z_{1})$ and $P(z_{2})$ 
 defined by eqn~(\ref{E-5A})) at two time moments.
 (a) - $t=$3~fs,  $\omega=$0.5~a.u.; 
 (b) - $t=$10~fs,  $\omega=$0.5~a.u.; 
  (c) - $t=$3~fs,  $\omega=$0.4~a.u.; 
 (d) - $t=$10~fs,  $\omega=$0.4~a.u..
  The other laser pulse parameters are: ${\cal E}_{0}=$0.02~a.u. and
$t_{p}=$5~fs.  Starting at $\pm 91$~a.u.
 the wave function is absorbed by an imaginary optical potential.
}
 \label{fig5}
 \end{figure*}
 The resonant excitation of atom~{\rm A} 
 at the laser carrier frequency $\omega =$0.5~a.u.
 provides an example of an {\it intuitive} case,
 when the ionization probability ${\rm I}_{\rm A}$ of the laser-excited atom~{\rm A}
 is large compared to the ionization probability ${\rm I}_{\rm B}$ of atom~{\rm B} 
 which is not excited by the laser field.
 The laser-driven dynamics of the extended H-H system in the case of 
 $\omega =$0.5~a.u. is presented in the left panel of Fig.~\ref{fig4} on a 10~fs timescale.
 For the sake of comparison, the right panel of Fig.~\ref{fig4} shows an example of
 a {\it counter-intuitive} case, $\omega =$0.4~a.u.,
 when the ionization probability ${\rm I}_{\rm B}$ of atom~{\rm B},
 which is not excited by the laser field, is substantially larger than 
 the ionization probability ${\rm I}_{\rm A}$ of the laser excited atom~{\rm A}.

 From Figs.~\ref{fig4}(a) and (d) one can see that the `atomic' energies $E_{\rm A}(t)$ 
 are controlled by the applied laser pulses:
 they increase in the first half of the pulses and decreases to the end of the pulses.
 In contrast, the `atomic' energies $E_{\rm B}(t)$ do not follow the applied laser pulses:
 they slowly increase in the second half of the pulses 
 and after the end of the pulses. 
 A similar behavior is found for `atomic' ionizations 
 ${\rm I}_{\rm A}$ and ${\rm I}_{\rm B}$ presented in Figs.~\ref{fig4}(b) and (e):
 while the laser-induced ionizations ${\rm I}_{\rm A}$ rise fast in the second half of the pulses,
 ionization probabilities ${\rm I}_{\rm B}$ start to rise only at the end of the pulses
 and sharply rise after the end of the pulses. 
 In the {\it counter-intuitive} case presented in Fig.~\ref{fig4}(e), 
 the ionization probability ${\rm I}_{\rm B}$ is almost two times larger than ${\rm I}_{\rm A}$.

 Taking into account that the laser pulses with narrow spatial envelopes
 [Fig.~\ref{fig1}(b)] excite only electron $e_{1}$ of atom~{\rm A} 
 in the vicinity of $z_{\rm A}=-50$~a.u.
 and do not affect electron $e_{2}$ of atom~{\rm B} in the vicinity of $z_{\rm B}=50$~a.u.,
 we can assume that the energy transfer from {\rm A} to {\rm B} [Figs.~\ref{fig4}(a) and (d)]
 and, especially, the time-delayed ionization of atom~{\rm B} 
 in the positive direction of the $z_{2}$-axis [Figs.~\ref{fig4}(b) and (e)], 
 occur entirely due to electron-proton attraction $V_{\rm e p}(z_{1},R)$ and
 the electron-electron repulsion $V_{\rm e e}(z_{1},z_{2})$
 in the vicinity of atom~{\rm B} at $z_{\rm B}=50$~a.u..

 In order to support these assumptions, we have calculated the electronic probabilities $P(z_{1})$ and $P(z_{2})$,
 defined by eqn~(\ref{E-5A}),  during the laser pulse  ($t=$3~fs) and after 5~fs free evolution of the H-H system
 after the end of the pulse ($t=$10~fs).  The results obtained are shown in Fig.~\ref{fig5} for the two excitation conditions of Fig. \ref{fig4}. First, let us focus on the resonant excitation case shown in panels (a) and (b) of Fig. \ref{fig5}. Because the laser pulse with the narrow spatial envelope
 excites only electron $e_{1}$ in the vicinity of $z_{\rm A}=-50$~a.u., 
the probability distribution $P(z_{1})$ during the pulse   [at $t=3$~fs,  Fig.~\ref{fig5}(a)]
 is broadened in comparison with the initial one [Fig.~\ref{fig1}(a)]
 in both $z_{1}<z_{\rm A}$ and $z_{1}>z_{\rm A}$ directions. 
 The laser-induced extension of $P(z_{1})$ into the domain of  $z_{1}<z_{\rm A}$ 
 gives rise to the ionization ${\rm I}_{\rm A}$ in the negative direction of the $z_{1}$-axis.
 In contrast, at $z_{1}>z_{\rm A}$, the laser-driven electron $e_{1}$ reaches
 the domain $z_{1}>0$ where it is attracted and accelerated by proton $p_{\rm B}$,
 localized at $z_{\rm B}=50$~a.u. [see Fig.~\ref{fig5}(b) for $t=10$~fs]. 
  \begin{figure*}[t]
 \begin{center}
 \includegraphics*[width=26pc]{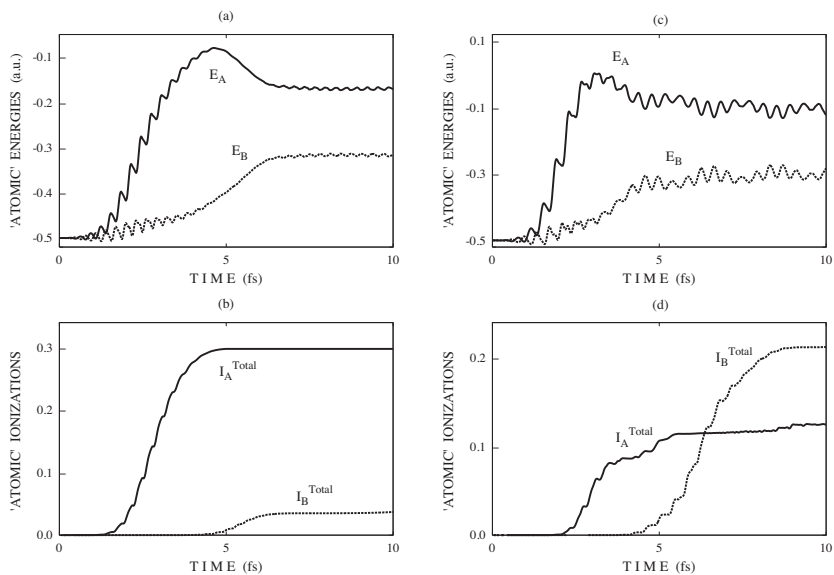}
 \end{center}
 \caption{Quantum dynamics of H-H excited from the entangled molecular singlet
 initial state by the narrowly shaped 5 fs laser pulse acting only on atom~{\rm A} 
 in the vicinity
 of $z_{\rm A}=-$50~a.u.: the {\it intuitive} atomic ionization
 (left panel, $\omega=$0.5~a.u.) and the {\it counter-intuitive}
 atomic ionization (left panel, $\omega=$0.4~a.u.).
 (a) and (c) - `atomic' energies $E_{\rm A}$ and $E_{\rm B}$
 defined by eqn~(\ref{E-EnAB-TS});
 (b) and (d) - total `atomic' ionizations defined by eqns.~(\ref{E-Flexes}).}
 \label{fig6}
 \end{figure*}
 %
 %
 
 The same qualitative behavior holds true for excitation with $\omega=0.4$ a.u. shown in Figs. \ref{fig5}(c) and (d). However, the details of the evolving wave packet are rather different giving rise to the different ionization probabilities. In particular we notice that due to the electron-proton Coulombic attraction, the probability $P(z_{1})$ has a local
 maximum at $z_{1}=z_{\rm B}$ [Figs.~\ref{fig5}(c) and (d)], 
 corresponding to laser induced electron transfer (LIET) of ref. 
 \citenum{Bandr:07prl.HpH2pBO}.
 Simultaneously, when electron $e_{1}$ approaches the domain of $z_{\rm B}=50$~a.u. and essentially localizes
 where the initial probability $P(z_{2})$ of electron $e_{2}$ has the global maximum [Fig.~\ref{fig1}(a)], 
 the electron-electron Coulombic repulsion becomes very strong. 
 Therefore, the probability $P(z_{2})$ is considerably extended into the domain of $z_{2}>z_{\rm B}$,
 giving rise to ionization ${\rm I}_{\rm B}$ in the positive direction of the $z_{2}$-axis. This effect is not so pronounced for resonant excitation where the wave packet describing electron $e_1$ is more delocalized in the region of atom B.
 
  Our numerical simulations performed with other laser carrier frequencies have shown 
 that the laser-driven dynamics of the electron probabilities $P(z_{1})$ and $P(z_{2})$
 and their evolution after the end of the laser pulse are similar to those
 presented in Fig.~\ref{fig5} for $\omega=$0.4~a.u..
 We can conclude therefore, that the underlying physical mechanisms of the energy
 transfer from the laser-excited atom~{\rm A} to atom~{\rm B} are
 the Coulombic attraction of the laser-driven electron $e_{1}$ of atom~{\rm A}
 by the proton $p_{\rm B}$ of atom~{\rm B} and a short-range Coulombic repulsion
 of electrons $e_{1}$ and $e_{2}$ when their wave functions significantly
 overlap in the domain of atom~{\rm B}.
\subsection{Entangled Initial States}  

 In the previous sections we have have presented results obtained 
 for the laser-driven dynamics
 of the extended H-H system of an "atomic origin" representing two distant
 H atoms. The initial state of such a system is an unentangled direct-product state. 
 It is instructive to compare this case to two other alternatives:
 excitation of the H-H system from entangled initial states, singlet and triplet,
 which represent an extended H-H system of a "molecular origin".
 
 For the sake of comparison to the {\it intuitive} and {\it counter-intuitive} cases
 of `atomic' ionizations, presented in Fig.~\ref{fig4} for the unentangled atomic initial state,
 we performed numerical simulations for the entangled molecular initial states, eq. (\ref{ini-ST}),
 by making use of the same laser fields as for the unentangled atomic initial state: 
 ${\cal E}_{0}=$0.02~a.u., $\omega=$0.5~a.u. and $\omega=$0.4~a.u., 
 with the narrow spatial envelope of the 5~fs laser pulse acting only on atom~{\rm A}.

 The laser-driven dynamics of extended H-H excited from  the entangled molecular singlet state is presented in Fig.~\ref{fig6}. Notice that we have found that the respective results for the initial triplet state are almost identical to those shown in Fig. \ref{fig6} and therefore are not presented.
 We can see that the time-dependent   `atomic' energies $E_{\rm A}(t)$ and $E_{\rm B}(t)$ 
 shown in Figs.~\ref{fig6}(a) and \ref{fig6}(c) are quite different from those 
 shown in Figs.~\ref{fig4}(a) and \ref{fig4}(d) for the unentangled atomic initial state.
 In contrast, the total time-dependent ionization probabilities 
 $I_{\rm A}^{\rm Total}$ and $I_{\rm B}^{\rm Total}$, 
 shown in Figs.~\ref{fig6}(b) and \ref{fig6}(d),
 are very similar (yet not identical) to the time-dependent ionization probabilities 
 $I_{\rm A}$ and $I_{\rm B}$ 
 presented in Figs.~\ref{fig4}(b) and \ref{fig4}(e) for the case of the unentangled atomic initial state.
 In particular, the {\it intuitive} ionization ($I_{\rm A} > I_{\rm B}$)
 takes place in both cases at $\omega=$0.5~a.u.,
 while the {\it counter-intuitive} ionization occurs in both cases at $\omega=$0.4~a.u..

 From the comparison of the results presented in Figs.~\ref{fig4} and \ref{fig6} the following
 observations can be made.

 (i) The overall energies transferred on a time scale of $t=10$~fs from atom~{\rm A}
 to atom~{\rm B} in the entangled molecular singlet and triplet states are almost 
 identical to each other and similar to that transferred from {\rm A} to {\rm B} 
 in the unentangled atomic state.

 (ii) In the case of the unentangled initial state, `atomic' energy $E_{\rm B}$
 starts to rise in the second half of the laser pulse and continues to rise
 after the end of the pulse, i.e., only when the laser-driven electron 
 of atom~{\rm A} approaches the domain of atom~{\rm B}. 
 In contrast, in the case of the entangled molecular initial states, the
 `atomic' energy of atom~{\rm B}, which is separated by 100~a.u.
 from the laser-excited atom~{\rm A},
 is controlled by the laser pulse acting only on atom~{\rm A}:
 the energy $E_{\rm B}(t)$ increases during the laser pulse and oscillates
 out-of-phase with $E_{\rm A}(t)$ during the laser pulse and after the end of the pulse.
 Such an entangled behavior of distant atoms during coherent excitation of one of them
 is important evidence for the possibility of long-range quantum communication.

 (iii) The time-dependent ionization probabilities are almost identical in all three
 cases of the initial states of the extended H-H system. 
 We can conclude therefore that the entanglement of the initial state of 
 the extended H-H system does not change its ionization probability as compared
 to the unentangled atomic initial state. 
 In particular, the {\it intuitive} atomic ionization remains {\it intuitive}
 and the {\it counter-intuitive} atomic ionization remains {\it counter-intuitive}.

 Note that similar observations have been made in our recent work 
 \cite{P-K-B:2011pra.H-H.extended}
 for the extended H-H system excited with a very different laser frequency
 of $\omega=$1.0~a.u.. 
 We can conclude therefore that observations (i)-(iii) are quite generic.

An interesting point concerns the question whether there is any difference at all
 in the laser-driven dynamics of the extended H-H system excited from an
entangled molecular initial states, singlet and triplet?  
%
 \begin{figure}[t]
 \begin{center}
 \includegraphics*[width=19pc]{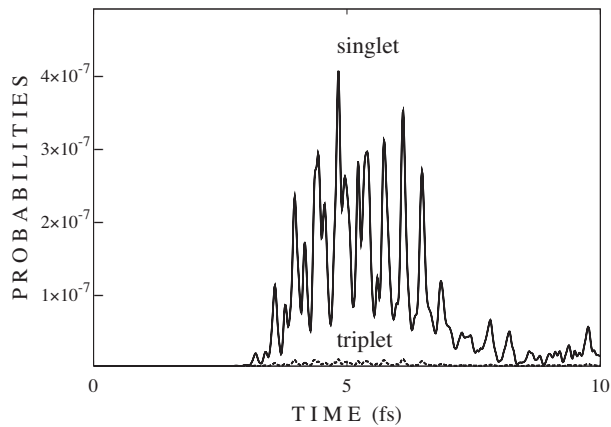}
 \end{center}
 \caption{The time-dependent probabilities to find two electrons 
 of the extended H-H system at the same point 
 for the singlet and triplet initial states.
 The probabilities are defined by eqn~(\ref{E-Pauli});
 the laser carrier frequency is $\omega=0.4$~a.u..}
 \label{fig7}
 \end{figure}
 In order to give a  quantitative answer to this question,
 we calculated the time-dependent integrated probability $P_{ee}(t)$  to find two electrons at the same point,
 \begin{equation}
 P_{ee}(t) = \int dR \int dz |\Psi (R,z,z,t)|^{2}.
 \label{E-Pauli}
 \end{equation}
 The results obtained at the laser carrier frequency $\omega=0.4$~a.u. 
 are presented in Fig.~\ref{fig7} 
 (in the case of $\omega=0.5$~a.u., both probabilities are about four times
 smaller as compared to those shown in Fig.~\ref{fig7}). 
 First we notice that even for the singlet state the probabilities $P_{ee}(t)$ of eqn~(\ref{E-Pauli})
 are very small. The respective probabilities for the triplet initial state are always by about two orders 
 of magnitude smaller in comparison to that of the singlet state.
 Therefore, the phase space available for electron-electron collisions
 in the triplet state is significantly smaller due to the antisymmetry 
 of the electron wave function in the triplet state
 in comparison to that in the singlet state. 
 Strictly speaking, $P_{ee}(t)$ should vanish for the initial triplet state and the finite but very small values shown in Fig.Ê
 \ref{fig7} reflect the numerical approximation.
 In order to study its influence  we have performed a numerical experiment, where we set $\Psi (R,z,z,t) = 0$ by hand at every time step. The result for the 'atomic' energies and ionization dynamics are virtually indistinguishable form those shown in Fig. \ref{fig6}.
 
 The results presented in Fig.~\ref{fig7} provide also a visualization of the electron-electron
 collisions for the singlet initial state of the extended H-H system. 
 Indeed, we can see that the probability to find both electrons of H-H at the same point
 in space reveals strong pulsations in the second half of the 5~fs laser pulse and
 after the end of the pulse, i.e., when the laser-driven electrons of atom~{\rm A}
 approach the domain of atom~{\rm B} where they interact with electrons of atom~{\rm B}. 
 What we  see, in fact, is how the kinetic energy of the laser-driven electrons
 of atom~{\rm A} competes with the potential energy of their repulsion  by the electrons of atom~{\rm B}.
%
 \section{Conclusion} 
 \label{sec:conc}
 %
 In the present work we have studied numerically the non-Born-Oppenheimer 
 quantum dynamics of two distant H atoms ({\rm A} and {\rm B})
 with an arbitrary but large initial internuclear
 separation of 100~a.u. (5.29~nm), referred to as the extended H-H system.
  Both unentangled atomic states and entangled molecular states, singlet and triplet,
 were used as initial states in our simulations. 
 
 First, the extended H-H system was assumed to be initially in an unentangled state,
 and only atom~{\rm A} was excited by the laser pulses with narrow spatial
 envelopes and various carrier frequencies.
 Here, we have found that
 an efficient energy transfer from atom~{\rm A} to atom~{\rm B}
 and ionization of atom~{\rm B} take place, 
 being induced by a short-range electron-electron repulsion 
 in the vicinity of atom~{\rm B} due to the preceding long-range
 transfer of the laser-driven electron $e_{1}$ from atom~{\rm A}
 enhanced by the Coulombic attraction and acceleration 
 of the laser-driven electron $e_{1}$ by proton $p_{\rm B}$ of atom~{\rm B}.
 Furthermore, we observed that in the low-frequency domain 
 $\omega \leq 0.6$~a.u.,
 EER-induced ionization of atom~{\rm B} 
 depends very strongly on the carrier frequency 
 of the laser pulse acting on atom~{\rm A}
 and can be both less efficient (an {\it intuitive} case)
 and more efficient (a {\it counter-intuitive} case)
 than the laser-induced ionization of atom~{\rm A}.

Both {\it intuitive} ($\omega=0.5$~a.u.) and {\it counter-intuitive} ($\omega=0.4$~a.u.)
 cases have been also studied for the entangled molecular states, singlet and triplet.
 We found that the time-dependent 
 `atomic' energies $E_{\rm A}(t)$ and $E_{\rm B}(t)$ 
 obtained for the entangled molecular states are quite different from those 
 obtained for the unentangled atomic initial state.
 In contrast, the total time-dependent ionization probabilities 
 $I_{\rm A}^{\rm Total}$ and $I_{\rm B}^{\rm Total}$ 
 for the entangled molecular initial states
 are very similar (yet not identical) to the time-dependent ionization probabilities 
 $I_{\rm A}$ and $I_{\rm B}$ 
 for the case of the unentangled atomic initial state.
 Moreover, both `atomic' energies and ionization probabilities in the extended
 H-H system excited from the initial triplet state
 are almost identical to those of the singlet initial state.
 We found that the physical reason for the indistinguishability of the laser-driven
 dynamics of the extended H-H system excited from the initial singlet and triplet state
 is a very small probability to find two electron of the extended H-H at the same point in space.

 We have shown in detail that in the case of the unentangled atomic initial state
 and a narrow spatial envelope of the applied laser field, 
 when only the electron initially belonging to atom~{\rm A} is excited by the laser field,
 the physical mechanisms of the energy transfer from atom~{\rm A} to atom~{\rm B} 
 and the ionization of atom~{\rm B} are as follows:
 (i) the Coulombic attraction of the laser-driven electron $e_{1}$ of atom ~{\rm A} 
 by proton $p_{\rm B}$ of atom~{\rm B},
 resulting, in the case of $\omega \ne 0.5$ a.u., in the formation of a narrow local maximum of the electronic wave function
 in the vicinity of proton $p_{\rm B}$,
 and (ii) the short-range Coulomb repulsion of electrons $e_{1}$ and $e_{2}$
 in the vicinity of proton $p_{\rm B}$ of atom~{\rm B},
 where their wave functions strongly overlap.
Due to the indistinguishability of the atomic ionization probabilities in extended H-H
 excited from the unentangled atomic state and from the entangled molecular states,
 both singlet and triplet, and on the basis of the results obtained in our recent work
 \cite{P-K-B:2011pra.H-H.extended},
 where the laser-driven dynamics of extended H-H has been studied 
 for the unentangled atomic state
 and for the entangled molecular singlet state 
 at  $\omega=1.0$~a.u.,
 we can conclude that the physical mechanisms of the energy transfer from atom~{\rm A}
 to atom~{\rm B} and the ionization of atom~{\rm B} in the entangled states are basically the same as those in the unentangled atomic state.
 
 Note finally that the narrow spatial envelope of the laser pulse, exciting only 
 one atom~{\rm A} of the extended H-H, used in the present work,
 is by no means unique, or obligatory for the
 long-range energy transfer to occur. 
 It has been shown in our recent work 
 \cite{P-K-B:2011pra.H-H.extended}
 that very similar results are obtained when atom~{\rm A}
 is excited predeominantly at the edge of the Gaussian spatial envelope of the laser pulse.
 The latter case is more likely to arise in an experiment.
  
%
\section*{Acknowledgement}
O.K. and G.K.P.  gratefully acknowledge stimulating discussions with Prof. D. Bauer (University of Rostock).
This work has been financially supported by the Deutsche Forschungsgemeinschaft 
 through the Sfb 652 (G.K.P., O.K.). A.D.B. thanks the Humboldt Foundation for the financial support through
 a Humboldt Research Award.

\footnotesize{
\providecommand*{\mcitethebibliography}{\thebibliography}
\csname @ifundefined\endcsname{endmcitethebibliography}
{\let\endmcitethebibliography\endthebibliography}{}

}
\end{document}